\documentclass[twocolumn]{article}
\usepackage{float}
\usepackage{graphicx}

\usepackage[utf8]{inputenc}
\usepackage[UKenglish]{babel}
\usepackage{multicol}
\usepackage[T1]{fontenc}     % improves font outcomes

\usepackage{amsmath}
\makeatletter
\renewcommand*\env@matrix[1][*\c@MaxMatrixCols c]{%
  \hskip -\arraycolsep
  \let\@ifnextchar\new@ifnextchar
  \array{#1}}
\makeatother

\usepackage{amsfonts}
\usepackage{amssymb}

\usepackage{lineno}
\usepackage[compress]{cite}
\usepackage{float}

\usepackage{graphicx}
\graphicspath{ {./figures/} } % There must not be space before and
\usepackage{caption} % Extend the formats for captions
\usepackage{subcaption}

\usepackage{array}
\usepackage{multirow}

\usepackage{url} % To add url-format to the document
\usepackage{mathtools} % More tools than those offered by "amsmath"
\usepackage{color} % Package to work with colours
\usepackage[usenames,dvipsnames,svgnames,table]{xcolor} % Add more colors
\usepackage{amsmath}

\usepackage[boxed, ruled, lined]{algorithm2e} % Support for algorithms
\usepackage{algpseudocode} % Required to use "algorithmic" environment
\usepackage{enumerate} % To enable using lower-case roman numbers in
\usepackage{framed}
\usepackage{listings}

\usepackage{hyperref}
\hypersetup{
  colorlinks=true,
  linkcolor=blue,
  filecolor=magenta,      
  urlcolor=cyan,
}

\usepackage[normalem]{ulem}
\numberwithin{equation}{section}

\usepackage{authblk}

\title{Immune cells interactions in the tumor microenvironment}
\author[1]{Mobina Tousian\thanks{m.tousianshandiz@uu.nl, mobinatousian@gmail.com}}
\author[2]{Christian Solis Calero
\thanks{csolisc@unmsm.edu.pe}}
\author[3]{Julio Cesar Perez Sansalvador\thanks{jcp.sansalvador@inaoe.mx}}
\affil[1]{Sharif University of Technology}
\affil[2]{Universidad Nacional Mayor de San Marcos}
\affil[3]{Instituto Nacional de Astrofísica, Óptica y Electrónica}
\date{}
\providecommand{\keywords}[1]{\textbf{\textit{Keywords---}} #1}

\begin{document}

\maketitle

\begin{abstract}
    % Something about tumour microenvironment
    
    % "**mobina: i add some parts of my draft for different topics
    % change whatever needed************************" "The study of the
    % microenvironment responses, effects, and gradients in which tumor
    % cells live; is one of the most important factors in the perusal of
    % tumor cell activities, proliferation, and annihilation. Immune
    % cells are the best-known foes fighting against tumor cells
    % provided by the host, but the accuracy of the microenvironment
    % complexity and biological characteristics are as important as
    % immune cell specifications; Since in biological environments
    % neglecting a factor or missing an interaction would cause serious
    % imbalances in chemical agents, resulting in fabled interactions
    % and an unreal perspective of the system response and its
    % biological conclusions."

 The tumor microenvironment (TME) plays a critical role in cancer cell proliferation, invasion, and resistance to therapy. A principal component of the TME is the tumor immune microenvironment (TIME), which includes various immune cells such as macrophages. Depending on the signals received from environmental elements like IL-4 or IFN-$\gamma$, macrophages can exhibit pro-inflammatory (M1) or anti-inflammatory (M2) phenotypes. This study uses an enhanced agent-based model to simulate interactions within the TIME, focusing on the dynamic behavior of macrophages. We examine the response of cancer cell populations to alterations in macrophages, categorized into three different behaviors: M0 (initial-inactive), M1 (immune-upholding), and M2 (immune-repressing), as well as environmental differentiations. The results highlight the significant impact of macrophage modulation on tumor proliferation and suggest potential therapeutic strategies targeting these immune cells.

\end{abstract}

\keywords{Tumor Immune Microenvironment, Agent-based model,
  Macrophages}

%%
%% Start line numbering here if you want
%%

\section{Introduction}

The tumor microenvironment (TME) is a complex, heterogeneous and dynamic system comprising cancer cells, immune cells, blood vessels, fibroblasts, signaling molecules, and the extracellular matrix (ECM)\cite{Norton:AR:2019}. Understanding the TME is crucial for advancing cancer research and developing effective therapeutic strategies\cite{Balkwill:AR:2012,Norton:AR:2019,Chamseddine:AR:2020,Bekisz:AR:2020,Kuznetsov:AR:2021}. Immune cells within the TME, particularly macrophages, play diverse roles in either supporting or inhibiting tumor progression. Macrophages can polarize into different phenotypes, with M1 macrophages exhibiting pro-inflammatory properties and M2 macrophages promoting immunosuppression and tissue repair\cite{Balkwill:AR:2012,Norton:AR:2019,Bekisz:AR:2020}. This study aims to explore the interactions between macrophages and cancer cells using an enhanced agent-based model to simulate the TIME.

\paragraph\noindent
The immune cells have the potential to recognize and eliminate cancer cells. However, the TME can create an immunosuppressive environment that inhibits immune cell activity and promotes tumor growth. Understanding the TME can help identify strategies to modulate the TME and enhance immune cell infiltration and function. Immunotherapy is a revolutionary approach that harnesses the power of the immune system to target and eliminate cancer cells. It involves the use of various strategies, such as immune checkpoint inhibitors, adoptive cell therapies (e.g., CAR-T cells), and cancer vaccines, to enhance the immune response against cancer. Immunotherapies aim to overcome the immunosuppressive TME and reinvigorate immune cell activity against cancer cells and tumor tissue.
\cite{Norton:AR:2019,Chamseddine:AR:2020,Bekisz:AR:2020,Kuznetsov:AR:2021,chen:AR:2017,binnewies:AR:2018}.

\paragraph\noindent
Macrophages are immune cells that can be recruited to the tumor microenvironment. TAMs play a complex role in cancer, as they can exhibit both pro-tumor and anti-tumor functions depending on their polarization state and the signals present in the tumor microenvironment
\cite{binnewies:AR:2018,gajewski:AR:2013}.

\paragraph\noindent
We study an enhanced multi-agent model of cancer treatment including macrophages, T-Cells, tumor cells, and microenvironment. we are able to predict behavior of the system in terms of populations of cells in early stages of body immune system noticing the presence of cancer cells, in response to alterations of macrophages, microenvironment and any cell types entangled in the process.  The microenvironment is the immediate environment that affects the performance of the system and consists of the cells, molecules, and structures that surround and support other cells and tissues.  The first basic feature of the network is the immune system's response to target and kill tumor cells. one of the most involved immune cells in the tumor-associated microenvironment(TME) is Macrophages. Macrophages can have different regarding their local microenvironment. Based on the signal macrophages would receive from their microenvironment they can have pro-inflammatory and immune-supporting properties or even immunosuppressive and wound-healing phenotypes\cite{cancers6031670}. In our model, we
categorize different macrophages’ behaviors into three phases: M0
(initial-inactive), M1(immune-upholding), and
M2(immune-repressing). Based on the type of local microenvironments cytokines M0 meets; it would turn into either M1 or M2 \cite{Metzcar:AR:2019}.

\section{Background}

Exploring the interactions among macrophages, cancer cells, and T cells is essential in understanding the adaptations that occur within the tumor microenvironment (TME). The normal immune response is often compromised in the TME, either through suppression or alteration into a pro-tumor state, leading to tumor proliferation. This study focuses on tumor-associated macrophages (TAMs), a prevalent and impactful type of immune cells within the TME. These macrophages not only possess cytotoxic abilities but also present antigens to T cells, which is part of their standard immune function and their behavior can significantly influence tumor progression and immune response dynamics \cite{Mantovani:AR:2002, Pollard:AR:2004, Noy:AR:2014}.

Interleukin-4 (IL-4) and Interferon-gamma (IFN-$\gamma$) are crucial cytokines that significantly influence the behavior of T cells, macrophages, and tumor cells within the tumor microenvironment.

IL-4 primarily promotes the differentiation of naive T-cells into Th2 cells, which are associated with humoral immunity and can help in the healing process by modulating immune responses. In macrophages, IL-4 also induces shift towards the M2 phenotype, which is generally associated with tissue repair and immuno-suppression, thus potentially supporting tumor growth and metastasis. The M2 macrophages, induced by IL-4, are involved in tumor progression and have a reduced capacity to combat cancer cells effectively\cite{Chen:AR:2023, Mills:AR:2000, Martinez:AR:2008}.

In contrast, IFN-$\gamma$, produced by Th1 cells, NK cells, and CD8+ T cells, which is a critical mediator of cellular immunity and has potential anti-tumor properties. It promotes the activation of macrophages to the M1 phenotype, which is more cytotoxic to tumor cells and capable of initiating robust immune responses against pathogens and tumors. IFN-$\gamma$ also enhances the ability of T-cells and other immune cells to recognize and destroy tumor cells through mechanisms like increasing the expression of major histocompatibility complex (MHC) molecules, which are vital for antigen presentation\cite{Castro:AR:2018, Schroder:AR:2004, Boehm:AR:1997}.

Both cytokines play pivotal roles in balancing immune responses: IL-4 often contributes to immunosuppression and tissue remodeling, while IFN-$\gamma$ is crucial for antiviral and antitumor responses, demonstrating how the immune system can be swayed in different directions based on the cytokine environment. This balance can determine whether the immune system supports or inhibits tumor growth, making these cytokines significant targets for cancer therapy and research into new treatments\cite{Chens:AR:2023,Castro:AR:2018}.

\section{Model}
\begin{figure}[h!]
\centering
	\includegraphics[width=\linewidth]{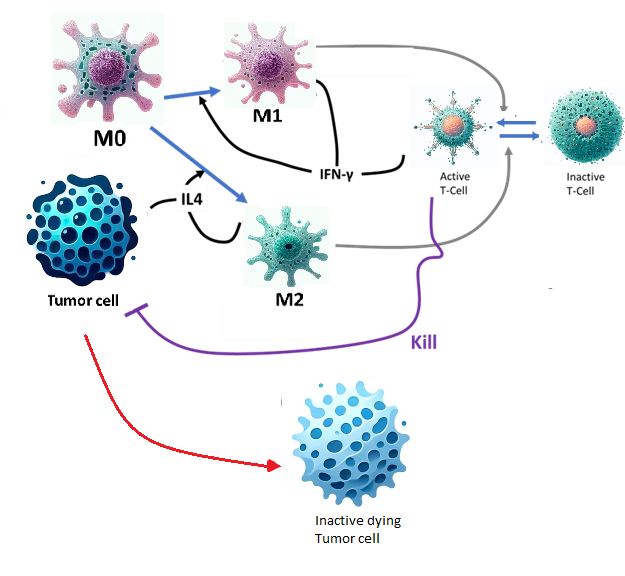}
\caption{The enhanced model}
\label{our_model}
\end{figure}

\begin{figure}[h!]
\centering
	\includegraphics[width=\linewidth]{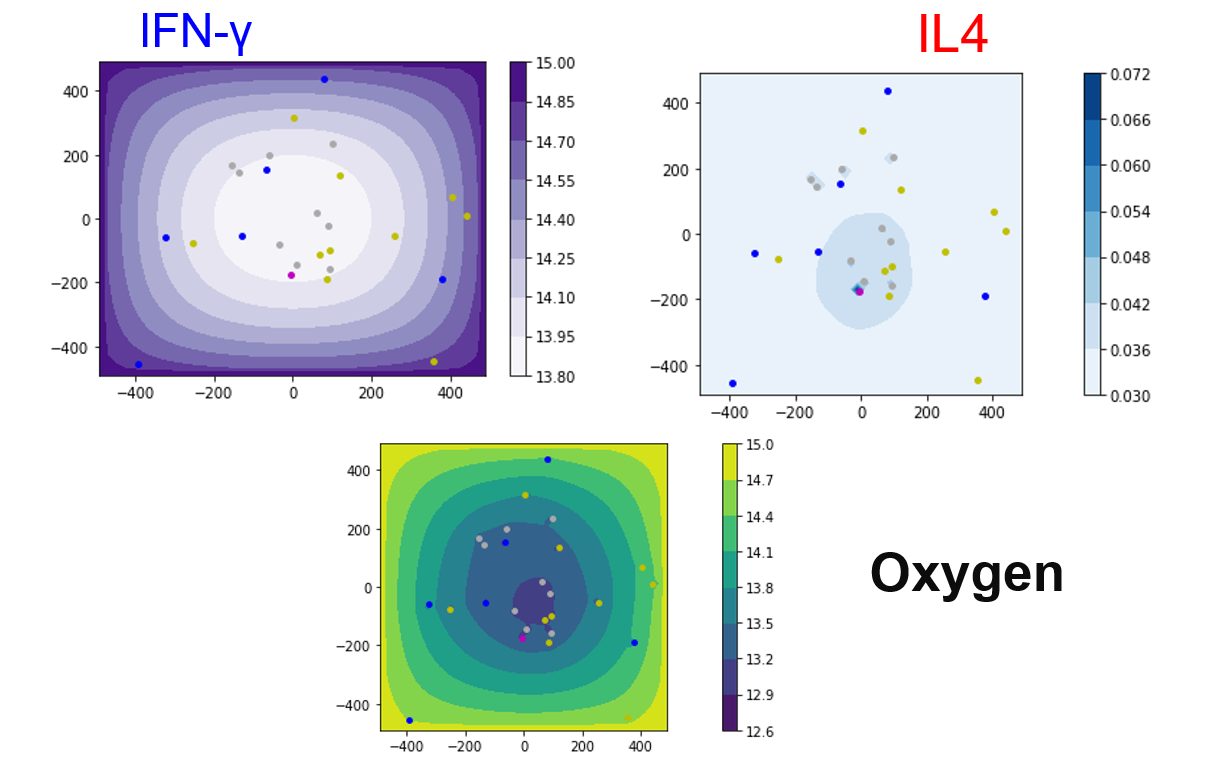}
\caption{Spatial distribution of microenvironment agents: IFN-$\gamma$ (top left), IL-4 (top right), and Oxygen (bottom). Color gradients represent concentration levels, influencing immune and tumor cell behaviors.}
\label{microenvironment}
\end{figure}

The original study by \cite{Cess:AR:2020} primarily focused on the interactions between macrophages, T cells, and tumor cells, emphasizing macrophage-based immunotherapies. Key features of the original model included Macrophage Differentiation and T Cell Activation.
Our model includes detailed dynamics of recruited T cells, focusing on their interactions with macrophages and tumor cells. Active T-Cells which are actively engaged in attacking tumor cells. Also states of T-cell exhaustion, reflecting the diminished ability of T-cells to respond to tumor cells over time due to chronic activation and inhibitory signals. Our model also includes more detailed cancer cell states. Cancer stem cells with the ability to self-renew and drive tumor growth shown as Tumor Cells in \ref{our_model} and Senescent Cells which are Cancer cells that have stopped dividing and can influence the tumor microenvironment through secretions.
When epithelial cancer cells die, they also release a substrate that promotes the differentiation of macrophages into the M1 phenotype, thereby enhancing the immune response against the tumor.
\paragraph\noindent
therefore the agents we used are Cancer cells, Macrophages, T-cells, and our microenvironment that consists of IFN-$\gamma$, IL-4, Oxygen, and Debris. For comparing and upgrading Colin's work; We ran simulations for both models. One consists of Tumor cells and T-cells and the other classifying T-cells into two subdivision of active and inactive and Tumor cells into two subdivisions of inactive mortal and operating active Tumor cells. 
Distribution ranges of key microenvironmental agents from some of our simulation domains are shown in \ref{microenvironment}. The contour plots represent the concentrations of IFN-$\gamma$, IL-4, and Oxygen. The concentrations are indicated by color gradients. These agents play crucial roles in regulating the behavior of immune cells and tumor cells. Concentrations of IFN-$\gamma$ and IL-4 influence macrophages' polarization and T-cell activation, while oxygen levels impact cell viability and proliferation.

\begin{figure}[h!]
\centering
	\includegraphics[width=\linewidth]{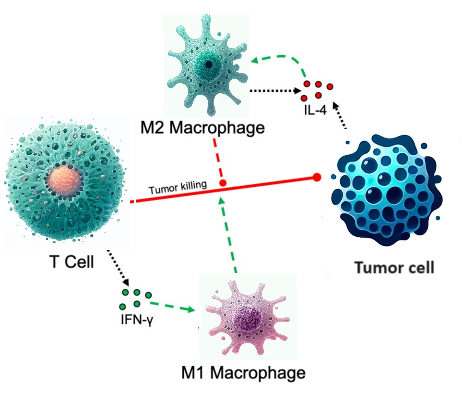}
\caption{colin's model \cite{Cess:AR:2020}.}
\label{colin_model}
\end{figure}

\subsection{Phenotypes}

\subsection*{Volume Phenotype}

The volume phenotype encompasses the dynamics of cellular volume regulation, which is a critical factor in cell physiology. It models how the cell's volume changes in response to environmental conditions and intrinsic cellular processes. The volume phenotype includes several sub-volumes, like the nuclear and cytoplasmic volumes, which are influenced by both solid and fluid states within the cell. Key parameters such as fluid change rates and biomass change rates govern how quickly the cell adapts its volume towards a target state. These parameters are expressed in the model through ordinary differential equations (ODEs) which simulate the rate of volume change over time.

\textbf{Formulas:}
The rate of change of fluid volume (\( \dot{V_F} \)) is modeled as:
\[
\dot{V_F} = r_F \times (V_F^* - V_F)
\]
where \( V_F^* \) is the target fluid volume, \( V_F \) is the current fluid volume, and \( r_F \) is the rate at which the volume changes towards the target.

Similarly, for nuclear (\( V_{NS} \)) and cytoplasmic solid volumes (\( V_{CS} \)), the rate of change is modeled by:
\[
\dot{V_{NS}} = r_N \times (V_{NS}^* - V_{NS})
\]
\[
\dot{V_{CS}} = r_C \times (V_{CS}^* - V_{CS})
\]
where \( V_{NS}^* \) and \( V_{CS}^* \) are the target nuclear and cytoplasmic solid volumes respectively.

\subsection*{T-cell phenotype}
\begin{itemize}
    \item A T-cell is in either an active or an inactive state.
    \item An inactive T-cell becomes active at contact with an M1 Macrophage
    \item An active T-cell becomes inactive at contact with an M2 Macrophage
    \item Active T-cells move by chemotaxis towards debris generated by dead cancer cells
    \item Active T-cells constantly secrete IFN-$\gamma$
    \item Active T-cell kill cancer cells.

\end{itemize}    
\subsection*{Macrophages phenotype}
\begin{itemize}
    \item There are three possible states for a Macrophage: M0, M1, or M2.
    \item An M0 Macrophage uptakes IFN-$\gamma$ and IL-4
    \item An M0 Macrophage may become an M1 Macrophage based on the uptaking of IFN-$\gamma$
    \item An M0 Macrophage may become an M2 Macrophage based on the uptaking of IL-4
    \item M1 Macrophages move by chemotaxis towards IFN-$\gamma$ secreted by active T-cells
    \item M2 Macrophages constantly secretes IL-4
    \item M2 Macrophages move by chemotaxis towards debris generated by dead cancer cells.\\

\noindent All cells are biased to move towards Oxygen and they will secret debris as they die.
\end{itemize}

\section{Methodology}
In our study, we utilized the PhysiCell simulation toolkit \cite{Ghaffarizadeh:AR:2018} to model the dynamics of cancer cell populations for different scenarios. The simulations were conducted to understand the interactions between cancer cells and their microenvironment, and to predict the population response to the immune system.

\subsection*{Computational Model}
PhysiCell provides a versatile framework for simulating cell populations through agent-based modeling. Each cell is treated as an independent agent characterized by specific biological properties defined by a set of differential equations and logical conditions.
\subsection*{Cell Cycle and Proliferation Model}

The cell cycle progression in our simulations was governed by a simplified version of a phase-based model, where each cell can exist in one of the following states: quiescent (G0), active proliferation (G1, S, G2, M), and apoptosis. The transition between these states is modeled using rate constants derived from experimental data provided by PhysiCell \cite{Ghaffarizadeh:AR:2018}:

\[
\text{Rate}_{\text{transition}} = \frac{1}{1 + \exp\left(-\left(\frac{\text{Signal} - \text{Threshold}}{\text{Sensitivity}}\right)\right)}
\]

This equation reflects the rate of transition from one cell cycle phase to another, driven by a signal relative to a threshold adjusted by sensitivity.

\subsection*{Microenvironment Interactions}
The secretion phenotype describes the cellular mechanisms involved in the secretion, uptake, and export of various biochemical substrates. This phenotype integrates seamlessly with the BioFVM framework to simulate the movement of chemical substrates within and between cells. The model accounts for the mass of substrates that cells either secrete into the extracellular matrix or uptake from it, modifying the concentration of these substrates locally. The secretion rate is influenced by the cell's volume, emphasizing how larger cells might have different metabolic rates compared to smaller ones.

Cells interact with their microenvironment, which includes gradients of nutrients and drugs. The diffusion and uptake of these substances are modeled using the following partial differential equations (PDEs):

\[
\frac{\partial C}{\partial t} = D \nabla^2 C - \lambda C + \sum_{i} \text{Source}_i - \sum_{i} \text{Uptake}_i
\]

Where \(C\) represents the concentration of a substrate (e.g., oxygen, drug), \(D\) is the diffusion coefficient, \(\lambda\) is the decay rate, and Source and Uptake terms represent the production and consumption by cells, respectively.

\subsection*{Mechanical Interactions}
The interactions between cells due to mechanical forces are crucial for understanding tissue morphology and tumor architecture. These forces include both adhesion and repulsion, modeled using linear spring equations that approximate the physical forces between cells. The force \(F\) exerted by one cell on another is calculated as:

\[
F = \begin{cases} 
k \cdot (d - r) & \text{if } r < d \\
0 & \text{otherwise}
\end{cases}
\]
Here, \(k\) represents the spring constant reflecting the strength of the mechanical interaction, \(r\) is the actual distance between the centers of two cells, and \(d\) is the equilibrium distance at which no force is exerted. When cells are closer than \(d\), they exert a repulsive force to return to this equilibrium state.

\subsection*{Simulation Execution}
The general time scales and constants are used from simulations which were conducted by PhysiCel and over a period representing several months to observe tumor growth and response to treatments, focusing on metrics such as survival rates, population diversity, and the development of drug resistance. We adjusted simulation parameters within biologically realistic limits to ensure the model accurately reflects observed experimental outcomes.
\paragraph\noindent
In this study, We ran the simulations for the early stages of the body's immune response. It is important to note that shorter simulation time steps can effectively capture the early dynamics and immediate responses of the tumor immune microenvironment (TIME) to various treatments and conditions. Studies have shown that initial interactions and responses within the TME can provide critical insights into tumor behavior and treatment efficacy. For instance, Norton et al. demonstrated that early-phase simulations are valuable for identifying key interactions and setting the stage for longer-term studies\cite{Norton:AR:2019}. Moreover, initial immune responses, such as macrophage polarization and cytokine interactions, often occur within the first hours of the immune system being triggered, making shorter simulations sufficient for certain research objectives\cite{Chens:AR:2023, Castro:AR:2018}; therefore the simulations are run for 48 hours with time step of a minute to also be able to study the microenvironments factor's map with changes that can be interpreted if monitored in a course of minutes. the initial number of cells were 300 for cancer cells, 50 for M0, T-Cell Inactive and T-Cell active.

\section{Discussion}

\begin{figure*}[h!]
  \centering
  \includegraphics[width=\textwidth]{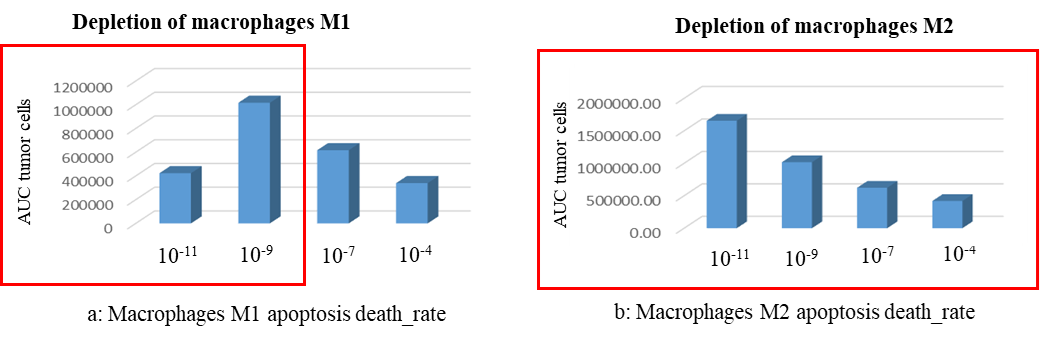}
  \caption{AUC Tumor cell values regarding different amounts of Macrophages a: M1 and b: M2 apoptosis death-rate}
  \label{AUC}
\end{figure*}

First we began to study the AUC of Tumor cells for the non-treated model of Colin\cite{Cess:AR:2020} as also shown in Figure(\ref{colin_model}). As we can see in Figure(\ref{AUC}), changing the apoptosis rate for M2 macrophages results in a decrease in the number of tumor cells over time on average. This decrease in tumor cells was expected because, in reality, active T-cells are responsible for the annihilation of Tumor cells. Macrophage M2, in contact with T-cells, converts active T-cells into inactive T-cells, and thus, M2 macrophages indirectly support tumor growth.

However, the situation is different for Macrophage M1. The initial assumption was that inactive T-cells by contact with macrophage M1, would turn into active T-cells; leading to a decrease in tumor cells. But as shown in Figure(\ref{AUC}), with higher depletion rates of macrophages M1, the average AUC of Tumor cells will increase initially but decreases afterward. This unexpected result highlights the complex dynamics within the tumor microenvironment. therefore we studied the effect of different apoptosis death rates of Macrophages on the AUC of all cell types with our new enhanced model shown as Figure(\ref{our_model}).

\begin{figure*}[h!]
\centering
	\includegraphics[width=\textwidth]{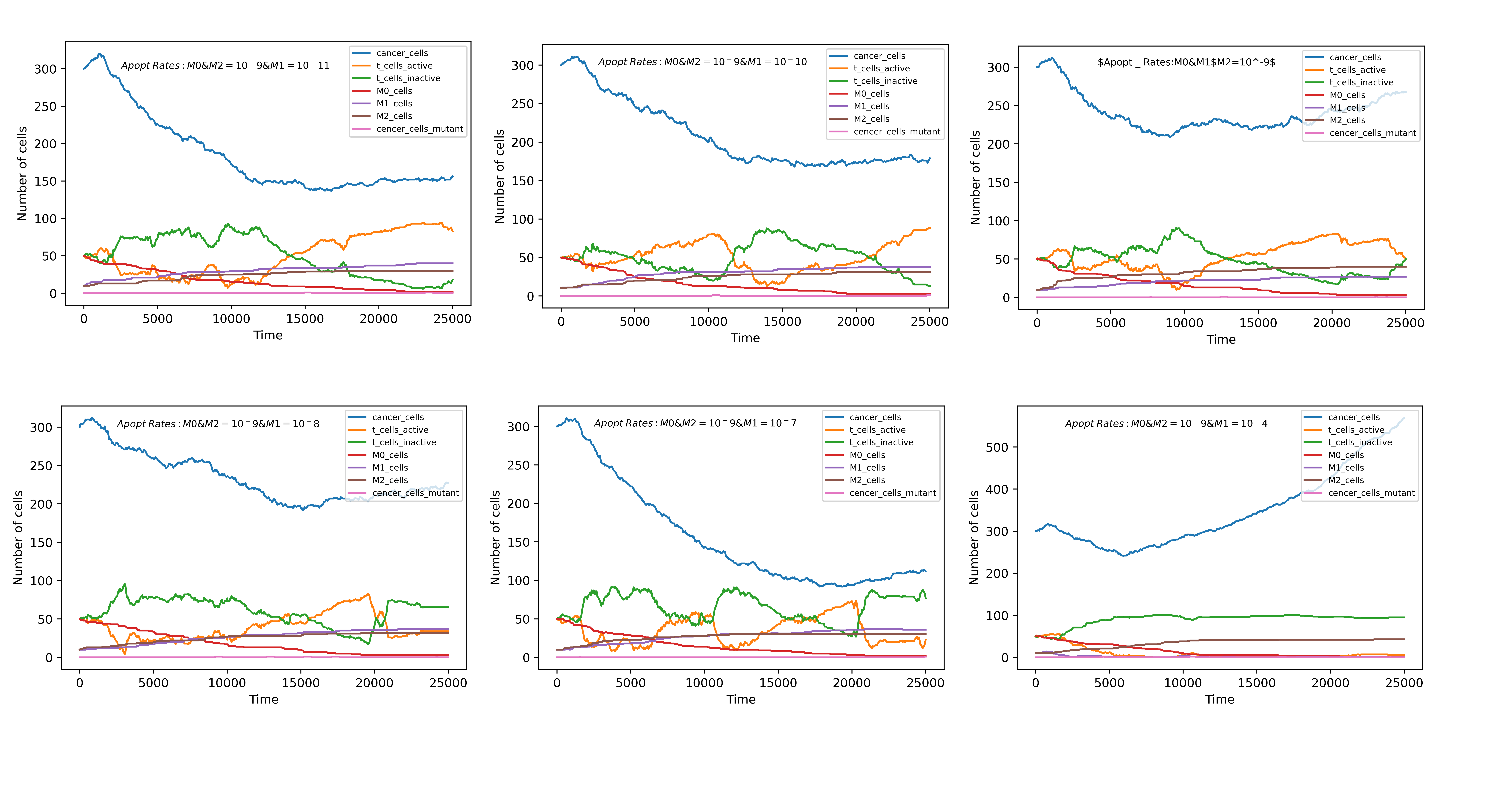}
\caption{AUC of Tumor cells regarding different amounts of Macrophages M1 apoptosis death-rate}
\label{M1_apoptosis_death_rates}
\end{figure*}

\begin{figure*}[h!]
\centering
	\includegraphics[width=\textwidth]{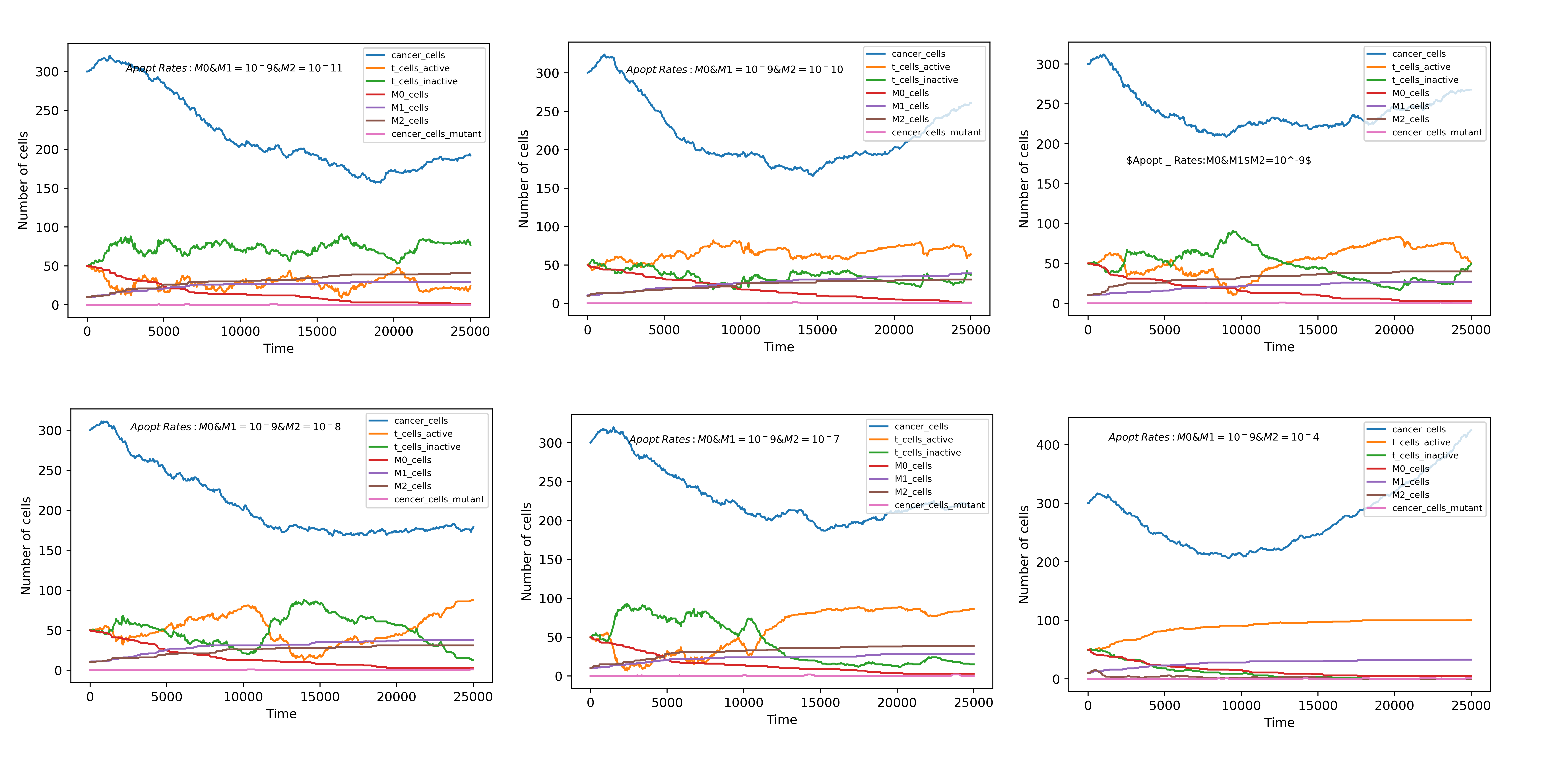}
\caption{AUC of Tumor cells regarding different amounts of Macrophages M2 apoptosis death-rate}
\label{M2_apoptosis_death_rates}
\end{figure*}

\begin{figure*}[htbp!]

\centering
	\includegraphics[width=\textwidth]{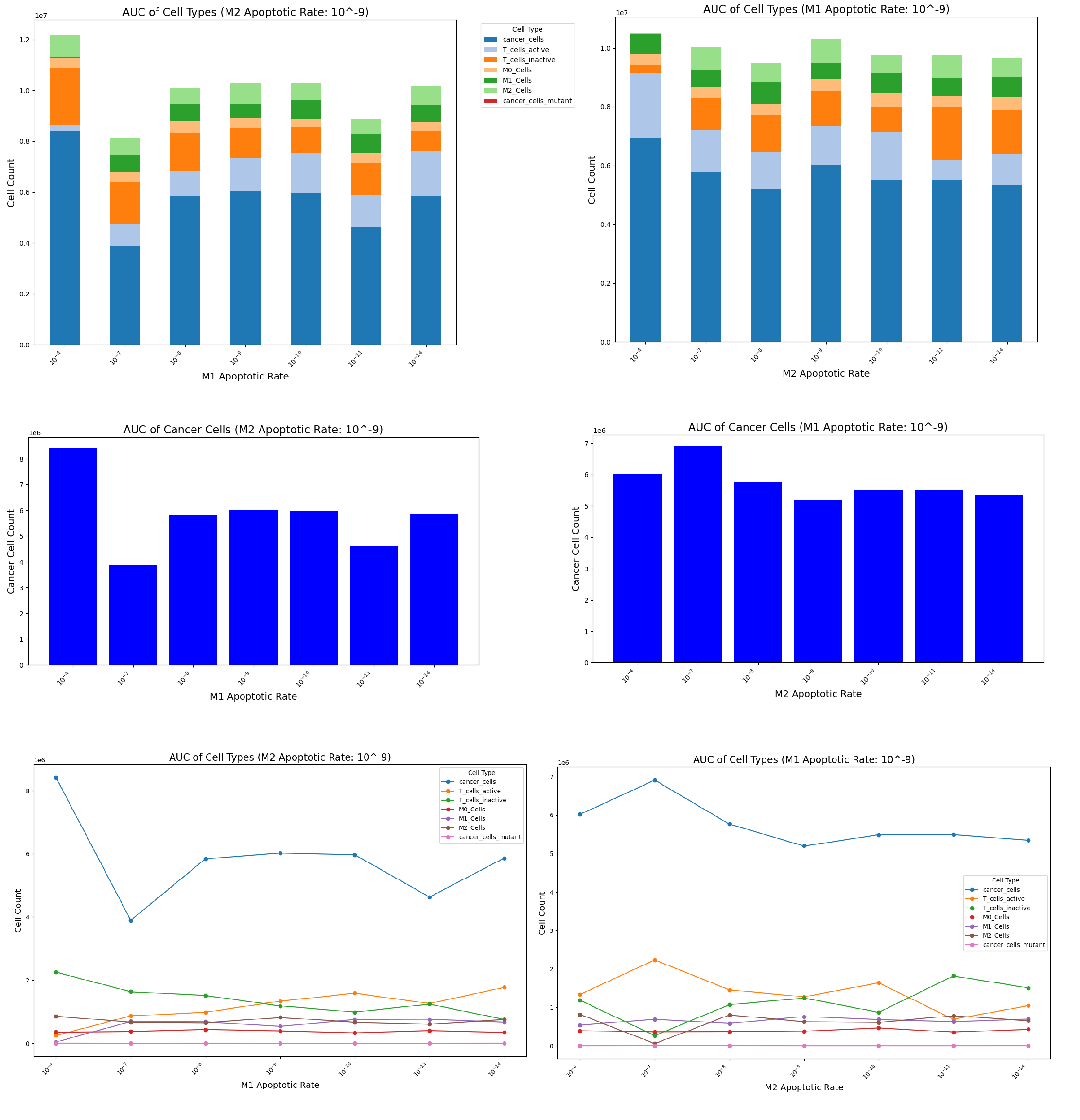}
\caption{AUC bar and line plot of all cells and Tumor cells regarding different amounts of Macrophages M1 and M2 apoptosis death-rate for the new enhanced model}
\label{M1_and_M2_apoptosismultibar}
\end{figure*}

Number of cells in some of simulations are plotted in Figure(\ref{M1_apoptosis_death_rates}) for different apoptosis rates of M1 and Figure(\ref{M2_apoptosis_death_rates}) regarding the changes in the apoptosis rates of M2.
As shown in Figure(\ref{M1_apoptosis_death_rates}), for all amounts of the death rates of Macrophage M1 and M2, we start with 300 cancer cells and then observe cell number variations in later time steps. For specific apoptosis rates like \( 10^-7 \) and \( 10^-14 \) for Macrophage M1, as shown in Figure(\ref{M1_apoptosis_death_rates}), an unpredictable intense decrease and increase in Tumor cells can be noticed. For a better understanding, AUC counts for all cells and cancer cells are shown in Figure(\ref{M1_and_M2_apoptosismultibar}).

M0 macrophages, which are undifferentiated, typically exhibit baseline apoptosis rates. Their apoptosis rates are relatively low, with half-lives generally reported to be around 24 to 48 hours under normal conditions. M1 macrophages are pro-inflammatory and more prone to apoptosis due to the harsh tumor microenvironment. The apoptosis rate for M1 macrophages is higher, with half-lives typically around 12 to 24 hours when exposed to strong inflammatory stimuli like IFN-$\gamma$. M2 macrophages, which are anti-inflammatory and tumor-supportive, have longer half-lives compared to M1 macrophages. They are more resistant to apoptosis and their half-lives can range from 24 to 72 hours, especially in a tumor-promoting environment where growth factors such as IL-4.

Typical reported values range from $1.67 \times 10^{-7}$ to $1.67 \times 10^{-6} \ \text{min}^{-1}$. M2 macrophages, which are anti-inflammatory and typically support tumor growth, exhibit lower apoptosis rates compared to M1 macrophages. Typical reported values range from $1.67 \times 10^{-8}$ to $1.67 \times 10^{-7} \ \text{min}^{-1}$. 

regardless of apoptosis rates of \( 10^-4 \) and \( 10^-14 \) as they would be extremely hard to be present in nature or genetically altered to be accessed; as it's shown in \ref{M1_and_M2_apoptosismultibar}, the natural choice of apoptosis rate for M1 to be around \(10^{-7}\) appears promising for reducing cancer cells, which can be easily achieved by changing experimental factors in cell cultures. and also for M2 macrophages, an apoptosis rate of \(10^{-9}\) seems effective for minimizing cancer cells. M2 is typically supporting tumor growth and it might cross the mind that higher M2 apoptosis rates would help the immune system. but regarding the complexity of the interactions of all the macrophages with microenvironment lower apoptosis rate such as \(10^{-9}\) for M2 is much more favorable to help the immune system overcome the rise in the Tumor cell populations in early stages of treatment.

Therefore the data suggest that maintaining M1 macrophage apoptosis rates around \( 10^-7 \) and M2 macrophage rates around \( 10^-9 \) could minimize tumor cell populations, supporting the therapeutic potential of modulating these rates.

Upon examining the pathway in the previous section, Figure(\ref{our_model}), we note that only active T-cells can annihilate tumor cells. Active and inactive T-cells are also only converted to each other via Macrophage M1 or M2. 
Therefore it can be assumed that the more of the M1 would be better for the reduction of cancer cells and the more of the M2 must do the reverse. But there are also modulations occurring for microenvirmental agents of IL4 and IFN-$\gamma$. 
So just like the inflection Point of \( 10^-9 \) for apoptosis rates differentiation, there can also be a breakpoint regarding the differentiation in uptake rates of micro-environments’ factors, IL4 and IFN-$\gamma$.

The uptake rates for cytokines like IFN-$\gamma$ and IL-4 can vary significantly depending on the type of cells and the conditions of the study. For IFN-$\gamma$ and IL-4, the uptake rate typically ranges from 1 to 10 molecules per minute per cell \cite{Janeway:IB:2001, Weinberg:BC:2013}.

\begin{figure*}[htbp!]

\centering
	\includegraphics[width=\textwidth]{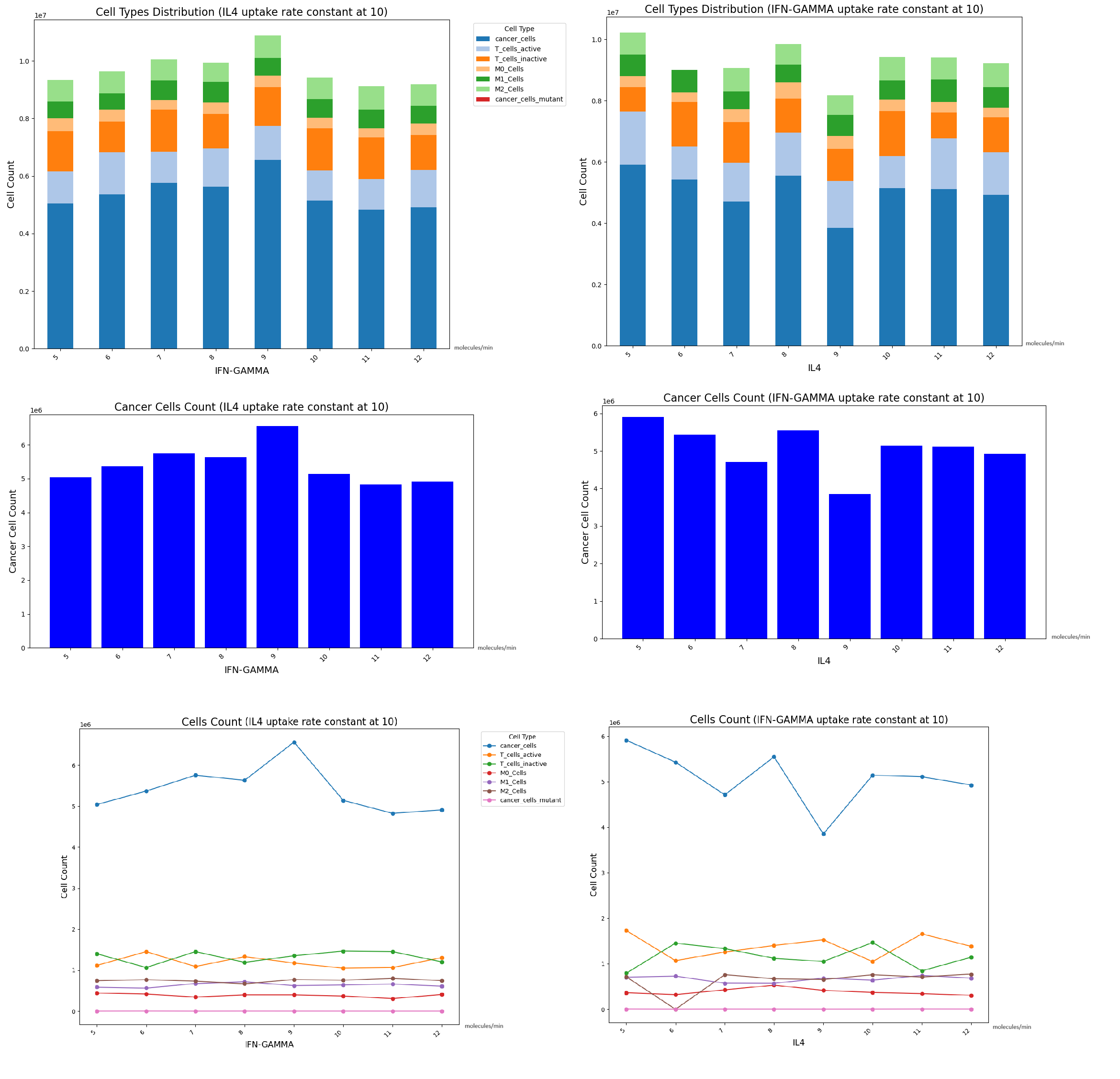}
\caption{AUC bar and line plot of all cells and Tumor cells regarding different amounts of IL4 and IFN-$\gamma$ uptake rates in the microenvironment for the new enhanced model}
\label{IL4_and_IFN_multibar}
\end{figure*}

In Figure(\ref{IL4_and_IFN_multibar}), we can see that the uptake rate of 9 (molecules per minute) would be ideal for IL4 and a disadvantage for IFN-$\gamma$.
\paragraph\noindent
Manipulating the uptake rates of IFN-$\gamma$ and IL-4 can be achieved through genetic engineering, pharmacological agents, and environmental manipulation. Techniques such as CRISPR/Cas9 can overexpress or knockdown cytokine receptors to enhance or manipulate uptake rates \cite{CRISPR2017, CytokineTherapy2019, Nanoparticle2021}.

\clearpage 
\section{Conclusion}
 Higher apoptosis rates of M2 macrophages correlate with a decrease in tumor cell numbers over time. This observation aligns with the understanding that M2 macrophages typically support tumor growth by suppressing the immune response. Their depletion, therefore, allows active T cells to more effectively target and destroy tumor cells. Contrary to initial assumptions, increasing the apoptosis rates of M1 macrophages initially leads to an increase in tumor cell numbers, followed by a decrease at higher apoptosis rates. This biphasic response suggests a complex interaction where moderate levels of M1 apoptosis might temporarily reduce immune support before higher depletion rates sufficiently disrupt tumor-supportive dynamics.
Both figures illustrate that varying the apoptosis rates of M1 and M2 macrophages significantly impacts tumor cell dynamics. Notably, apoptosis rates of \( 10^-4 \) and \( 10^-7 \)  for M1 macrophages show unpredictable fluctuations, highlighting the non-linear and sensitive nature of these interactions.
Our enhanced model demonstrates the significant role of macrophage behavior and cytokine interactions in tumor dynamics. The findings underscore the importance of fine-tuning macrophage apoptosis rates and cytokine uptake rates to optimize therapeutic outcomes. Future studies should focus on validating these results in experimental settings and exploring additional factors that influence the TIME.
\section*{Acknowledgments}
We would like to express our deepest gratitude to Dr. Reza Ejtehadi from Sharif University of Technology for his invaluable guidance and support throughout this study. We are truly grateful for his encouragement and for providing us with the environment necessary to carry out this research.
I would like to acknowledge Dr. Reza Ejtehadi from Sharif University of Technology for his supervision during my study. I am truly grateful for his encouragement to carry out my research, his overall guidance and the environment he provided were instrumental in my academic development.

% ========================================================================
% Customising bibliography [BibTex] ---------- BEGIN ---------------------
% ========================================================================
\onecolumn

\bibliography{biblio}
\bibliographystyle{ieeetr}
%\bibliographystyle{plain}
%\bibliographystyle{alpha}
%\bibliographystyle{siam}
%\bibliography{biblio}
% ========================================================================
% Customising bibliography [BibTex] ---------- END -----------------------
% ===========dead========================================================

\end{document}